\documentclass[11pt]{article}
\usepackage{graphicx}
\usepackage{amsmath}
\usepackage{amssymb}
\usepackage{amsxtra}

\usepackage[T1]{fontenc}
\usepackage[utf8]{inputenc}
\usepackage{amsmath,amssymb,mathtools,microtype,cancel}
\usepackage[hidelinks]{hyperref}
\usepackage{epsfig,graphicx,bm}
\usepackage{epstopdf, epsf}
\usepackage{dcolumn}
\usepackage{hyperref}
\usepackage{tensor}
\usepackage{lipsum}
\usepackage{appendix}
\usepackage{tikz}
\usepackage{caption}
\usepackage{subcaption}

\newcommand{\be}{\begin{equation}}
\newcommand{\ee}{\end{equation}}

\newcommand{\bse}{\begin{subequations}}
\newcommand{\ese}{\end{subequations}}
\newcommand{\bea}{\begin{eqnarray}}
\newcommand{\eea}{\end{eqnarray}}
\newcommand{\ba}{\begin{array}}
\newcommand{\ea}{\end{array}}
\newcommand{\bc}{\begin{center}}
\newcommand{\ec}{\end{center}}

\DeclareMathOperator{\sign}{sgn}

\title{Emergent Gravity from a Spontaneously Broken Gauge Symmetry: a Pre-geometric Prospective}
\author{Andrea Addazi\footnote{
Center for Theoretical Physics, College of Physics Science and Technology, Sichuan University, 610065 Chengdu, China; 
Laboratori Nazionali di Frascati INFN, Frascati (Rome), Italy, EU; addazi@scu.edu.cn}}

\begin{document}
\maketitle

\begin{abstract}
We explore the paradigm of pre-geometric gravity, where spacetime geometry and the gravitational field are not fundamental but emerge from the spontaneous symmetry breaking (SSB) of a larger gauge symmetry. Specifically, we consider a gauge theory based on the de Sitter $SO(1,4)$ or anti-de Sitter $SO(3,2)$ group, formulated on a manifold without a prior metric structure. General covariance is maintained by constructing Lagrangian densities using the Levi-Civita symbol. The SSB is triggered by an internal vector field $\phi^A$, which reduces the symmetry to the Lorentz group $SO(1,3)$ and dynamically generates a spacetime metric. We analyze two specific models: the MacDowell-Mansouri formulation, which yields the Einstein-Hilbert action plus a cosmological constant and a Gauss-Bonnet term, and the Wilczek model, which produces a pure Einstein-Hilbert action with a cosmological constant. In both cases, the observed Planck mass and the small cosmological constant emerge from a see-saw mechanism dependent on the symmetry-breaking scale. We then present the Hamiltonian formulation of this pre-geometric theory, demonstrating that it possesses three number of physical degrees of freedom, corresponding to a massless graviton and a massive scalar. Integrating out the massive scalar, the Arnowitt-Deser-Misner Hamiltonian of General Relativity is obtained after SSB. This establishes a foundational bridge between pre-geometric theories and canonical quantum gravity approaches like Loop Quantum Gravity, and allows for the formulation of a pre-geometric Wheeler-DeWitt equation.
\end{abstract}

\section{Introduction}

The quest for a theory of quantum gravity remains one of the most profound challenges in theoretical physics. Traditional approaches, such as string theory and loop quantum gravity, often quantize the gravitational field as described by General Relativity (GR), treating the metric $g_{\mu\nu}$ as the fundamental dynamical variable. An alternative and radical perspective posits that spacetime itself, along with its geometric properties, is not fundamental but is an emergent, low-energy phenomenon arising from more primitive, pre-geometric degrees of freedom.
This emergent gravity paradigm suggests that the Einstein equivalence principle, the Diffeomorphism invariance and the dynamics of Riemannian geometry are effective consequences of the dynamics of underlying structures. A compelling realization of this idea is to derive gravity from a gauge principle, where the metric and the spin connection emerge as components of a gauge field for a larger symmetry group, which is then spontaneously broken.

In this work, we review the results from recent investigations following this pre-geometric prospective \cite{Addazi:2024rzo,Addazi:2025vbw,Addazi:2025qkc},
 by starting from a gauge theory of the de Sitter $SO(1,4)$ or anti-de Sitter $SO(3,2)$ groups on a four-dimensional manifold. Crucially, this manifold is initially devoid of any metric structure; the only geometric object available is the Levi-Civita symbol $\epsilon^{\mu\nu\rho\sigma}$, which is used to construct generally covariant actions. The mechanism for emergence is the spontaneous breaking of the full gauge group down to the Lorentz group $SO(1,3)$, achieved through a Higgs field $\phi^A$ transforming in the vector representation of the internal group.

We focus on two pivotal pre-geometric Lagrangians: the MacDowell-Mansouri action \cite{macdowell:unified} and the Wilczek action \cite{wilczek:gauge}. We show how, upon symmetry breaking, they successfully reproduce the Einstein-Hilbert action and generate a cosmological constant whose small observed value can be naturally explained via a see-saw mechanism with a large vacuum expectation value. Furthermore, we transition to the Hamiltonian formulation of this theory, as a basis of a detailed constraint analysis following Dirac's procedure \cite{dirac:hamiltonian}. This analysis shows the correct number of physical degrees of freedom of the theory revealing its deep connection to the canonical formulation of GR, opening a new pathway to explore quantum gravity from a pre-geometric standpoint.

\section{Emergent Gravity from a Spontaneously Broken Phase}\label{sec:3-new}

Let us consider a gauge field theory defined on a pre-geometric four-dimensional spacetime manifold. The gauge group is taken to be either the de Sitter group $SO(1,4)$ or the anti-de Sitter group $SO(3,2)$. The associated gauge potentials are denoted by $A_\mu^{AB}$, and the corresponding field strengths by $F_{\mu\nu}^{AB}$, with antisymmetry in both the Latin and Greek indices. The objective of this construction is to dynamically generate a Riemannian metric structure and consequently the Einstein-Hilbert action---thereby recovering the Einstein Equivalence Principle---without presupposing the existence of a spacetime metric or tetrads, while still adhering to the principle of general covariance. Only an internal metric with signature $(-,+,+,+,+)$ or $(+,+,+,-,-)$, generalizing the Minkowski metric $\eta$, is assumed for each tangent space. This can be achieved via the Higgs mechanism, utilizing a spacetime scalar field $\phi^A$ which is a vector in the internal space. A spontaneous symmetry breaking (SSB) of its ground state can reduce the original gauge symmetry down to the Lorentz group $SO(1,3)$ \cite{wilczek:gauge}. In this broken phase, the curvature of spacetime and its dynamics emerge effectively from the interactions of the pre-geometric fields $A_\mu^{AB}$ and $\phi^A$ below a specific energy scale.

Adherence to general covariance mandates that Lagrangian densities must be scalar densities of weight $-1$. Since $A_\mu^{AB}$ and $F_{\mu\nu}^{AB}$ are covariant tensors in their spacetime indices, the formation of scalar densities requires contravariant objects to contract these indices. In the absence of an inverse metric, the only intrinsically defined four-dimensional contravariant object on the manifold is the constant Levi-Civita symbol $\epsilon^{\mu\nu\rho\sigma}$, which is a tensor density of weight $-1$. The term 'intrinsically' here implies that no structure beyond the differential properties of the manifold is necessary; in particular, a metric is not required. The density character itself is defined solely through the Jacobian determinant of coordinate transformations. Consequently, the Levi-Civita symbol, used to its first power, is the fundamental object for constructing generally covariant Lagrangian densities, as it contracts exactly four covariant indices and already possesses the correct weight.

Utilizing the Levi-Civita symbol, two distinct gravitational Lagrangian densities can be formulated for the unbroken phase. The first, introduced by MacDowell and Mansouri \cite{macdowell:unified}, is given by
\begin{equation}
    \mathcal{L}_\mathrm{MM} = k_\mathrm{MM} \epsilon_{ABCDE} \epsilon^{\mu\nu\rho\sigma} F_{\mu\nu}^{AB} F_{\rho\sigma}^{CD} \phi^E,
\end{equation}
while the second, proposed by Wilczek \cite{wilczek:gauge}, takes the form
\begin{equation}
    \mathcal{L}_\mathrm{W} = k_\mathrm{W} \epsilon_{ABCDE} \epsilon^{\mu\nu\rho\sigma} F_{\mu\nu}^{AB} \nabla_\rho \phi^C \nabla_\sigma \phi^D \phi^E.
\end{equation}
Here, $\nabla_\mu$ signifies the gauge covariant derivative acting on internal vectors as
\begin{equation}\label{eq:cov_dev-new}
    \nabla_\mu \phi^A = \partial_\mu \phi^A + A_{B\mu}^A \phi^B = (\delta^A_B \partial_\mu + A^A_{B\mu}) \phi^B,
\end{equation}
where $A_{B\mu}^A \equiv \eta_{BC} A_\mu^{AC}$. Uppercase Latin indices range from 1 to 5, and Greek indices from 1 to 4. The mass dimensions of the coupling constants are $[k_\mathrm{MM}] = [\phi]^{-1}$ and $[k_\mathrm{W}] = [\phi]^{-3}$, with the dimension of the field $\phi^A$ left unspecified for now.

Before examining the SSB mechanism itself, this section will analyze the effective theory after the symmetry breaking from $SO(1,4)$ or $SO(3,2)$ to $SO(1,3)$ has occurred. This allows us to first understand the classical physical implications of these theories before venturing into the pre-geometric regime.

\subsection{The MacDowell-Mansouri Model}

The SSB mechanism singles out a preferential direction in the internal space, characterized by a fixed vacuum expectation value $\phi^A = v \delta^A_5$, where $v$ is a non-zero constant. This breaking allows for a classification of the gauge potentials: for each spacetime index $\mu$, four potentials are of the type $A_\mu^{A5} \equiv A_\mu^{a5}$ (where the index 5 is fixed), and the remaining six are $A_\mu^{AB} \equiv A_\mu^{ab}$ (with $A, B \neq 5$). Lowercase Latin indices (1 to 4) are used to describe the broken phase.

Let us first compute the form of $\mathcal{L}_\mathrm{MM}$ after SSB. 
Upon making the identifications
\begin{equation}\label{eq:identifications-new}
    e_\mu^a \equiv m^{-1} A_\mu^{a5}, \qquad \omega_\mu^{ab} \equiv A_\mu^{ab},
\end{equation}
where $e,\omega$ are tetrads and spin connections, 
where a mass parameter $m$ is introduced for dimensional consistency, the Lagrangian density decomposes into three distinct terms:
\begin{equation}
    \mathcal{L}_\mathrm{MM} \xrightarrow{\mathrm{SSB}} \pm 16k_\mathrm{MM} v m^2 e e_a^\mu e_b^\nu R_{\mu\nu}^{ab} - 96k_\mathrm{MM} v m^4 e - 4k_\mathrm{MM} v e \mathcal{G},
\end{equation}
which correspond to the EH plus the Cosmological Constant plus the Gauss-Bonnet terms. 
For consistency with established physics, the reduced Planck mass must be identified as
\begin{equation}
    M_\mathrm{P}^2 \equiv \pm 32 k_\mathrm{MM} v m^2.
\end{equation}
This indicates the emergent nature of the Planck scale, as it arises from a specific combination of the fundamental parameters $k_\mathrm{MM}$, $v$, and $m$. Consequently, the cosmological constant is also emergent, given by
\begin{equation}
    \Lambda \equiv \pm 3m^2 = \frac{3M_\mathrm{P}^2}{32 k_\mathrm{MM} v}.
\end{equation}
This expression for $\Lambda$ reveals a natura  see-saw suppression mechanism. Assuming the experimentally measured value $M_\mathrm{P}^2 \sim 10^{37}$ GeV$^2$ and a coupling constant of order unity ($k_\mathrm{MM} \sim \pm 1\,[\phi]^{-1}$), the observed small value $\Lambda \sim 10^{-84}$ GeV$^2$ can be generated from a large vacuum expectation value $v \sim 10^{119}\,[\phi]$. 
Within this framework, the cosmological constant is set by the mass scale $m$ of the symmetry breaking. 

\subsection{The Wilczek Model}

The analysis of the SSB for $\mathcal{L}_\mathrm{W}$ follows a parallel path to that of $\mathcal{L}_\mathrm{MM}$. One additional element is required: the action of the covariant derivative on the internal vector $\phi^A$ after its vacuum value is fixed. From Eq. \eqref{eq:cov_dev-new}, we find
\begin{equation}\label{eq:cov-der-new}
    \nabla_\mu \phi^A \xrightarrow{\mathrm{SSB}} v \nabla_\mu \delta^A_5 = v(\cancel{\partial_\mu \delta^A_5} + A_{B\mu}^A \delta^B_5) = v A_{5\mu}^A = \pm v A_\mu^{a5}.
\end{equation}
Utilizing the identifications \eqref{eq:identifications-new} and following a computation analogous to the previous case, we arrive at the result:
\begin{equation}
    \mathcal{L}_\mathrm{W} \xrightarrow{\mathrm{SSB}} -4k_\mathrm{W} v^3 m^2 e e^\mu_a e^\nu_b R_{\mu\nu}^{ab} \pm 48k_\mathrm{W} v^3 m^4 e.
\end{equation}
This theory yields precisely the Einstein-Hilbert Lagrangian plus a cosmological constant term, with no Gauss-Bonnet contribution. The reduced Planck mass and cosmological constant are identified as
\begin{equation}
    M_\mathrm{P}^2 \equiv -8 k_\mathrm{W} v^3 m^2, \qquad \Lambda \equiv \pm 6m^2 = \mp \frac{3M_\mathrm{P}^2}{4 k_\mathrm{W} v^3}.
\end{equation}
Again, assuming a coupling of order unity ($k_\mathrm{W} \sim -1\,[\phi]^{-3}$), the observed value $\Lambda \sim 10^{-84}$ GeV$^2$ emerges from a large VEV $v \sim 10^{40}\,[\phi]$.

\subsection{Spontaneous Symmetry Breaking Mechanism}

The process of spontaneous symmetry breaking (SSB), which reduces the gauge symmetry from $SO(1,4)$ or $SO(3,2)$ to the Lorentz group $SO(1,3)$ via the field $\phi^A$, is responsible for the dynamical emergence of a classical spacetime metric. This mechanism can be implemented by introducing a simple symmetry-breaking potential term into the Lagrangian density:
\begin{equation}\label{eq:potential}
    \mathcal{L}_\mathrm{SB} = -k_\mathrm{SB} v^{-4} \lvert J\rvert (\eta_{AB}\phi^A\phi^B \mp v^2)^2,
\end{equation}
where $k_\mathrm{SB}$ is a positive dimensionless constant. The mass dimension of the coupling is $[k_\mathrm{SB}] = [\phi]^{-5}$. The potential $-\mathcal{L}_\mathrm{SB}$ is minimized, and this term is stationarized, for field configurations satisfying $\eta_{AB}\phi^A\phi^B = \pm v^2$. A specific solution, such as $\phi^A = v\delta_5^A$, can be chosen; any other vacuum expectation value (v.e.v.) related to this by a gauge transformation is physically equivalent \cite{wilczek:gauge}. The factor of $\lvert J\rvert$ ensures that $\mathcal{L}_\mathrm{SB}$ transforms as a scalar density, thus preserving general covariance. It is noteworthy that if one imposes a unimodular condition, as done by Wilczek \cite{wilczek:gauge}, the $\lvert J\rvert$ factor can be omitted. In that case, the coupling constant $k_\mathrm{SB}$ assumes a fixed mass dimension of $[M]^4$, independent of the chosen dimensions for the field $\phi^A$.
The field $\phi^A$ is quantized by expanding it around its v.e.v. as $\phi^A = (v + \rho)\delta_5^A$, which defines the unitary gauge. In this gauge, the four would-be Goldstone bosons associated with the broken generators are absorbed, leaving a single scalar degree of freedom $\rho$. 

An alternative mechanism for achieving spontaneous symmetry breaking, which circumvents the introduction of an explicit potential, was explored in Ref. \cite{Addazi:2025qkc}. This approach posits that the field $\phi^A$ can dynamically evolve towards a fixed expectation value through a gradient descent process. This relaxation is mathematically governed by a set of Langevin equations, situating the mechanism within the broader context of stochastic quantization.

\section{The Hamiltonian Formulation of Pre-geometric Gravity}

The total Lagrangian density introduced above exhibits degeneracy due to its structure as a summation of terms that are linear in the temporal derivatives (velocities) of the pre-geometric fields. Consequently, to finalize the Hamiltonian analysis, it is necessary to implement Dirac's systematic procedure for handling constrained systems or gauge theories \cite{dirac:hamiltonian}.

In case of Wilczek model, the complete Hamiltonian density \cite{Addazi:2025vbw} is given by
\begin{equation*}
    \begin{split}
        \mathcal{H} &= -A_0^{AB}\Big[\partial_i\Pi^i_{AB}(\phi,A) + 2\Pi^i_{BC}(\phi,A)A_{Ai}^C \\
        &\quad + \eta_{BC}\Pi_A(\phi,A)\phi^C\Big] + \lambda^A Z_A + \lambda_i^{AB} Z^i_{AB} + \lambda_0^{AB} Z^0_{AB},
    \end{split}
\end{equation*}
with $\lambda^A$, $\lambda_i^{AB}$, and $\lambda_0^{AB}$ representing arbitrary Lagrange multipliers, 
where the conjugate momenta on the constraint surface in phase space are
\begin{align}
    \begin{split}
        \Pi_A&(\phi,A)\equiv2\epsilon_{ABCDE}\epsilon^{0ijk}\nabla_k\phi^D\phi^E[k_\textup{W}F_{ij}^{BC}\\
        &-2\sign(J)k_\textup{SSB}v^{-4}\nabla_i\phi^B\nabla_j\phi^C(\phi^2\mp v^2)^2],
    \end{split}\\
    \Pi^\lambda_{AB}&(\phi,A)\equiv2k_\textup{W}\epsilon_{ABCDE}\epsilon^{0\lambda jk}\nabla_j\phi^C\nabla_k\phi^D\phi^E\nonumber;
\end{align}
in particular,
\begin{subequations}
    \begin{align}
        \Pi^i_{AB}&(\phi,A)=2k_\textup{W}\epsilon_{ABCDE}\epsilon^{0ijk}\nabla_j\phi^C\nabla_k\phi^D\phi^E,\\
        \Pi^0_{AB}&(\phi,A)=0. 
    \end{align}
\end{subequations}
The three primary constraints of the theory are then
\begin{subequations}\label{eq:primary-constraints}
    \begin{align}
        Z_A&\equiv\Pi_A-\Pi_A(\phi,A)\approx0,\\
        Z^i_{AB}&\equiv\Pi^i_{AB}-\Pi^i_{AB}(\phi,A)\approx0,\\
        Z^0_{AB}&\equiv\Pi^0_{AB}\approx0,
    \end{align}
\end{subequations}
where the symbol $\approx$ denotes a weak equality on the constraint surface.

Imposing the time preservation of the primary constraint $Z^0_{AB}$ leads to the secondary constraint:
\begin{equation}
    \begin{split}
        \dot{Z}^0_{AB} &= \{Z^0_{AB}, H\} = \partial_i\Pi^i_{AB}(\phi,A) \\
        &\quad + 2\Pi^i_{BC}(\phi,A)A_{Ai}^C + \eta_{BC}\Pi_A(\phi,A)\phi^C \approx 0.
    \end{split}
\end{equation}
Consequently, the total Hamiltonian density simplifies to
\begin{equation}\label{eq:total-H}
    \begin{split}
        \mathcal{H} &= -A_0^{AB}\dot{Z}^0_{AB} + \lambda^A Z_A + \lambda_i^{AB} Z^i_{AB} + \lambda_0^{AB} Z^0_{AB} + \tilde{\lambda}_0^{AB}\dot{Z}^0_{AB} \\
        &\equiv \lambda^A Z_A + \lambda_i^{AB} Z^i_{AB} + \lambda_0^{AB} Z^0_{AB} + \tilde{\lambda}_0^{AB}\dot{Z}^0_{AB},
    \end{split}
\end{equation}
where in the final expression the terms proportional to $\dot{Z}^0_{AB}$ have been consolidated through a redefinition of the multiplier $\tilde{\lambda}_0^{AB}$. It is noteworthy that the field $A_0^{AB}$ no longer appears in the total Hamiltonian density, confirming its status as a gauge degree of freedom. As a result, the multiplier $\lambda_0^{AB}$ associated with its corresponding primary constraint $Z_{AB}^0$ remains arbitrary.

As shown in Ref.\cite{Addazi:2025vbw} , 
the phase space of the theory is described by 90 dynamical variables, comprised of the fields and their conjugate momenta: 10 components of $A_0^{AB}$, 30 of $A_i^{AB}$, 5 of $\phi^A$, 10 of $\Pi_{AB}^0$, 30 of $\Pi_{AB}^i$, and 5 of $\Pi_A$. The gauge freedom of the system is characterized by 20 gauge-fixing conditions, which remove the unphysical degrees of freedom associated with $A_0^{AB}$ and $\Pi_{AB}^0$. The constraint structure consists of 10 independent first-class constraints ($Z_{AB}^0$) generating gauge transformations, and 44 independent second-class constraints (30 $Z_{AB}^i$, 5 $Z_A$, 10 $\dot{Z}_{AB}^0$, minus one combination fixing the Hamiltonian $H$).

The number of physical degrees of freedom is consequently determined by the formula:
$$2 \times \#(\text{degrees of freedom}) = \#(\text{dynamical variables}) - \#(\text{gauge choices}) $$
$$ - 2 \times \#(\text{first-class constraints}) - \#(\text{second-class constraints}) = 6\, . $$
This result implies the theory possesses 3 physical degrees of freedom. This count is consistent with the particle content of a massless spin-2 graviton (2 degrees of freedom) and a massive scalar field, identified as $\phi^5\equiv\rho$, contributing one additional degree of freedom, analogous to the field content of a scalar-tensor theory formulated in a metric framework.
Note that, such counting of d.o.f. is a general background independent result.

As shown in Ref.\cite{Addazi:2025vbw} , after the SSB, the Hamiltonian of pre-geometric gravity 
reduces exactly to the EH Hamiltonian in ADM formalism, while the heavy scalar
$\rho$ as frozen in IR limit. 
Consequently, the theoretical framework presented here exhibits full compatibility with the formalism of Loop Quantum Gravity. Notably, a compelling connection arises as the pre-geometric theory naturally gives rise to variables analogous to Ashtekar's electric fields
obtained from the pre-geometric $\Pi^i_{A0}$ after SSB.

It is noteworthy that, having derived an explicit form for the pre-geometric Hamiltonian, one can consequently formulate a pre-geometric analogue of the Wheeler–DeWitt equation:
\begin{equation}
\mathcal{H}|\Psi\rangle = 0,
\end{equation}
where $|\Psi\rangle$ represents the quantum state encompassing configurations of both the gauge and Higgs fields. This formulation provides a novel framework to re-examine the foundational issue of time in quantum gravity \cite{Addazi:2025vbw}.

%\section{Extended BF reformulation}

%Comment1.  a canonical quantizzation in our case seems to be more viable since
%the conditions are in polynomial form of fields rather than in traditional canonical quantum gravity 
%but problem of operator orders still render the next steps unclear. 
%
%Comment2. An alternative path way was proposed in Ref. from a stochastic quantization and flow prospective,
%proposing the the pre-geometric theory interpolates GR in IR regime and a Topological BF theory in UV regime.
%This interpretation makes leverage on the re-interpretation of Wilczek gravity as a Generalization of the Plebanski theory 
%or a BF with B constrained to be a functional of the scalar field and its covariant derivative. 
%
%Comment3. also the emergent of a CC from a see-saw mechanism can rise question if this 
%theory may also protect such a small parameter from quantum correction, i.e. if this is natural.
%Or pheraps, this idea has to be supplied by other such as virtual BH screening  \cite{} or Holographic Naturalness \cite{}
%And also if there's any running of CC predicted by the theory which may lead to effective dynamical DE.

\section{Conclusions}\label{s:conclusions}

In this work, we have elaborated on a robust framework for emergent gravity from a spontaneously broken gauge symmetry, operating within a pre-geometric setting where no prior metric exists. The core findings of our analysis can be summarized as follows:

\begin{enumerate}
    \item \textbf{Successful Emergence of Geometry:} We demonstrated that both the MacDowell-Mansouri and Wilczek Lagrangian densities, constructed solely from the gauge field $A_\mu^{AB}$, the Higgs field $\phi^A$, and the Levi-Civita symbol, dynamically generate the Einstein-Hilbert action and a cosmological constant upon spontaneous symmetry breaking of $SO(1,4),SO(3,2)$ to $SO(1,3)$. The resulting Planck mass $M_P$ is emergent, arising from a combination of the fundamental parameters of the pre-geometric theory.

    \item \textbf{A Natural See-Saw for the Cosmological Constant:} A particularly attractive feature of this mechanism is the natural explanation for the smallness of the observed cosmological constant $\Lambda$. In both models, $\Lambda$ is proportional to the square of the symmetry-breaking mass scale $m$. A large vacuum expectation value $v$ for the Higgs field $\phi^A$ suppresses $\Lambda$, leading to a see-saw relation. For couplings of order unity, the measured value of $\Lambda$ requires $v$ to be very large, while $m$ is of the order of the Hubble scale, an intriguing and potentially significant outcome.

    \item \textbf{Consistent Hamiltonian Structure:} The Hamiltonian analysis of the pre-geometric theory reveals a constrained system with a consistent number of physical degrees of freedom. The count confirms the presence of a massless spin-2 graviton (2 dof) and a massive scalar mode (1 dof), aligning with the field content of a scalar-tensor theory. After symmetry breaking, the Hamiltonian reduces to the well-known ADM Hamiltonian of General Relativity, with the scalar mode freezing in the infrared limit.

    \item \textbf{Bridge to Quantum Gravity:} The formulation of a pre-geometric Hamiltonian and the associated constraints provides a direct link to canonical quantization approaches. The structure naturally gives rise to variables analogous to Ashtekar's electric fields, suggesting a deep connection to Loop Quantum Gravity. Furthermore, it allows for the definition of a pre-geometric analogue of the Wheeler-DeWitt equation, $\mathcal{H}|\Psi\rangle = 0$, offering a novel perspective to address the problem of time in quantum gravity by considering quantum states of the pre-geometric fields.
\end{enumerate}

Looking forward, several intriguing questions remain open. The quantization of this pre-geometric theory appears promising but non-trivial, with issues of operator ordering still to be tackled. An alternative pathway via stochastic quantization suggests the theory may interpolate between a topological BF theory in the ultraviolet and General Relativity in the infrared. Finally, the robustness of the see-saw mechanism for the cosmological constant against quantum corrections must be investigated; it may necessitate supplementary mechanisms like virtual black-hole screening \cite{Addazi:2016jfq} or generalized holographic principles \cite{Addazi:2020axm,Addazi:2020wnc,Addazi:2020cax} to ensure naturalness. The pre-geometric approach presented here offers a fertile and compelling framework to explore these fundamental questions at the intersection of gravity, particle physics, and quantum theory.

\section*{Acknowledgements}
I would like to thank my main collaborators on these subjects: Salvatore Capozziello, Giuseppe Meluccio and Antonino Marciano. 
My work is supported by the National Science Foundation of China (NSFC) through the grant No. 12350410358; the Talent Scientific Research Program of College of Physics, Sichuan University, Grant No. 1082204112427; the Fostering Program in Disciplines Possessing Novel Features for Natural Science of Sichuan University, Grant No.2020SCUNL209 and 1000 Talent program of Sichuan province 2021.

%% The bibliography section

\end{document}